\documentclass[twocolumn,showpacs,preprintnumbers,amsmath,amssymb,prl,superscriptaddress]{revtex4}
\usepackage{mathrsfs}
\usepackage{graphicx}
\usepackage{dcolumn}
\usepackage{bm}
\usepackage{amsmath}
\usepackage{amsfonts}

\begin{document}

\title{Topological Magnetic Insulators with Corundum Structure}
\author{Jing Wang}
\affiliation{Department of Physics, Tsinghua University, Beijing
100084, China} \affiliation{Department of Physics, McCullough
Building, Stanford University, Stanford, CA 94305-4045}
\author{Rundong Li}
\affiliation{Department of Physics, McCullough Building, Stanford
University, Stanford, CA 94305-4045}
\author{Shou-Cheng Zhang}
\affiliation{Department of Physics, McCullough Building, Stanford
University, Stanford, CA 94305-4045}
\author{Xiao-Liang Qi}
\affiliation{Microsoft
Research, Station Q, Elings Hall, University of California, Santa
Barbara, CA 93106}
\affiliation{Department of Physics, McCullough Building, Stanford
University, Stanford, CA 94305-4045}

\date{\today}

\begin{abstract}
Topological insulators are new states of quantum matter in which
surface states residing in the bulk insulating gap
are protected by time-reversal symmetry. When a proper kind of
antiferromagnetic long range order is established in a topological
insulator, the system supports axionic excitations. In this paper,
 we study theoretically the electronic states in a transition metal
oxide of corundum structure, in which both spin-orbit
interaction and electron-electron interaction play crucial roles. A
tight-binding model analysis predicts that materials with this
structure can be strong topological insulator. Because of the
electron correlation, an antiferromagnetic order may develop, giving rise to a topological magnetic insulator phase with axionic excitations.
\end{abstract}


\maketitle

The discovery of time reversal invariant topological insulator has
attracted great attention in condensed matter
physics~\cite{bernevig2006d,koenig2007,fu2007a,hsieh2008,zhang2009,xia2009,chen2009,qi2010,hasan2010}.
With time-reversal symmetry broken on the surface, the
electromagnetic response of three dimensional (3D) insulators are
described by the topological $\theta$ term of the form
$S_{\theta}=\frac{\theta}{2\pi}\frac{\alpha}{2\pi}\int
d^3xdt\mathbf{E\cdot B}$ together with the ordinary Maxwell terms,
where $\mathbf{E}$ and $\mathbf{B}$ are the conventional
electromagnetic field insides the insulator, $\alpha=e^2/\hbar c$ is
the fine structure constant, and $\theta$ is the dimensionless
pseudoscalar parameter describing the insulator, which refers to
``axion'' field in axion electrodynamics~\cite{wilczek1987}. For a
system without boundary, all the physical quantities are invariant
if $\theta$ is shifted by integer multiple of $2\pi$. Therefor all
time reversal invariant insulator fall into two distinct classes
described by either $\theta=0$ (trivial insulator) or $\theta=\pi$
(topological insulator)\cite{qi2008}. Such a universal value of
$\theta=\pi$ in topological insulators leads to magneto-electric
effect with an universal coefficient, which has several unique
experimental consequences such as a topological contribution to the
Faraday rotation or Kerr rotation\cite{qi2008,tse2010,maciejko2010},
and the image monopole induced by an electron\cite{qi2009}. $\theta$
has an explicitly microscopic expression of the momentum space
Chern-Simons form which depends on the band structure of the
insulator~\cite{qi2008}
\begin{equation}
\theta=\frac{1}{4\pi}\int
d^3k\epsilon^{ijk}\mathrm{Tr}\left[A_i\partial_jA_k+i\frac{2}{3}A_iA_jA_k\right],
\end{equation}
where $A_i^{\mu\nu}(\mathbf{k})=-i\langle u_{\mu}|\partial/\partial
k_i|u_{\nu}\rangle$ is the momentum space non-abelian gauge field,
with $|u_{\mu}\rangle$, $|u_{\nu}\rangle$ referring to the Bloch
wavefunction of occupied bands.

If strong electron correlation exists in a topological insulator, a
long-range antiferromagnetic(AFM) order can be established under low
enough temperature. Since the AFM order breaks time-reversal
symmetry spontaneously, $\theta$ can deviate from $\pi$, and also
becomes a {\it dynamical} field which has fluctuations associated
with some spin collective modes. The spin collective mode inducing
fluctuations of $\theta$ are thus coupled to the photons by
$\theta{\bf E\cdot B}$ term, which means they become ``axions" in
the term used in high energy physics. Such a nonconventional
antiferromagnetic insulator supporting axion excitations is proposed
as topological magnetic insulator (TMI)~\cite{Li2010}. Due to its
coupling to photons, the axion field hybridizes with photons,
leading to axion polariton, with a polariton gap tunable by an
external magnetic field. Thus such a material can be used as a novel
type of optical modulator to control the transmission of light
through the material.

To realize the TMI phase, we need both the nontrivial topology of
the electron bands and strong electron correlation. The materials
with electrons in 4d or 5d-orbital can have both strong spin-orbit
coupling (SOC) and strong interaction, which is ideal for this
purpose. Recently, models for topological insulators with strong
electron correlation have been
proposed\cite{raghu2008,Nagaosa2009,guo2009,balents2010}, also
first principle calculations show topological phases exist in
thallium-based III-V-VI$_2$ ternary
chalcogenides\cite{yan2010,chen2010} as well as ternary heusler
compounds which contain the rare earth element Ln, where additional
properties ranging from superconductivity to magnetism and
heavy-fermion behavior can be realized\cite{chadov2010,lin2010}. In
this Letter, we study theoretically the transition metal oxide
$ABO_3$ of corundum structure with $B$ and $A$ standing for some
transition metals such as Fe, Ti, Ru, Rh, Ir, Os,
etc~\cite{Imada1998}. A tight binding model is obtained by using
point group symmetry of this structure, from which we find a
topological magnetic insulator phase with certain SOC strength and
electron-electron interaction.

\begin{figure}[t]
\centering
\includegraphics[width=3.0in,clip=true]{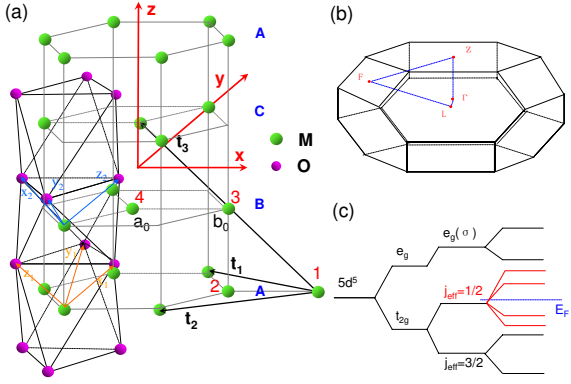}
\caption{(color online) (a) Corundum crystal structure with three
primitive lattice vector denoted as $\vec{t}_{1,2,3}$. Each
transition metal ion M (M$=$Ir,~Os,~etc.) (green large circles) is
surrounded by oxygen octahedron (red small circles). Each unit cell
has four M atoms, denoted as $1,2,3,4$. The space group is
$R\bar{3}c$. (b) Brillouin zone for corundum structure with space
group $R\bar{3}c$. The four inequivalent time-reversal invariant
points are $\Gamma(000)$, $F(\pi00)$, $L(\pi\pi\pi)$ and $Z(00\pi)$.
(c) Schematic crystal field splitting of $5d$ level in corundum
structure. We are interested in the half filling case with effective
angular momentum $j_{\rm eff}=1/2$.}
\label{fig1}
\end{figure}

The corundum structure is shown in Fig.~\ref{fig1}(a). Each
transition metal atom is surrounded by oxygen octahedron, and the d
orbitals are split by the octahedral crystalline field into doublet
$e_g(x^2-y^2,3z^2-r^2)$ and triplet $t_{2g}(xy,yz,zx)$ orbitals (See
Fig.~\ref{fig1}(c)). We will neglect small distortion of the
oxygen octahedra which may lead to minor corrections to electronic
structure~\cite{Nebenzahl1971}. The energy of $t_{2g}$ stays lower
with respect to $e_g$, because the latter point towards the
negatively charged oxygens. The SOC is effective in $t_{2g}$
orbitals and negligible in $e_g$ orbitals. Including the SOC,
$t_{2g}$ splits into total angular momentum $j_{\mathrm{eff}}=3/2$
and $j_{\mathrm{eff}}=1/2$. We focus on those materials where the
Fermi level lies completely in the $j_{\mathrm{eff}}=1/2$ sub-bands.
For example, the ions $\mathrm{Ir^{4+}}$, $\mathrm{Os}^{3+}$,
$\mathrm{Ru}^{3+}$ etc with five $d$-electrons satisfy this
requirement\cite{Prewitt1969}.

To obtain the electron dynamics in this system, we start by a
symmetry analysis to the corundum structure. The space group of this structure is
$D_{3d}^5(R\bar{3}c)$ with four atoms in each unit cell. It has a
trigonal axis (three-fold rotation symmetry $C_3$) defined by $z$
axis, a binary axis (two-fold rotation symmetry $C_2$), defined by
$y$ axis, and inversion symmetry with the inversion center at the
middle of the two neighbor transition metal atoms. The primitive
lattice vectors $\vec{t}_{1,2,3}$ and primitive unit cells are shown
in Fig.~\ref{fig1}(a), where each unit cell consists of four
transition metal atoms denoted as 1,2,3,4. Since the O $p$-level
$\epsilon_p$ are far away from the fermi level, we can consider a
model describing only $d$-electrons, with the hopping mediated by
the oxygen $p$-orbitals. The model is generally written as
\begin{equation}\label{H}
\mathcal{H}_0=-\sum\limits_{\langle
ij\rangle}\left[d_i^{\dagger}t_{ij}d_j+h.c.\right]+
\sum\limits_{\langle\langle
ij\rangle\rangle}\left[d_i^{\dagger}\hat{t}^{\prime}_{ij}d_j+h.c.\right],
\end{equation}
where $\langle ij\rangle$ and $\langle\langle ij\rangle\rangle$
denote the nearest-neighbor (NN) and next-nearest-neighbor (NNN)
sites, respectively, and the hopping terms $t_{ij}$ and
$\hat{t}_{ij}'$ are in general $2\times 2$ matrices. The form of the
parameters $t_{ij}, \hat{t}_{ij}'$ can be simplified by symmetry
considerations. Due to space limitation, we will only present the
result of the symmetry analysis. The NN transfer integral $t_{ij}$
are real and spin independent, with two independent parameters, the
intra-plane hopping $t$ and the inter-plane hopping $t_{\perp}$.
$t=(pd\pi)^2[(pp\sigma)+3(pp\pi)]/3(\epsilon_d-\epsilon_p)^2$~\cite{Nagaosa2009},
where $(pd\pi)$, $(pp\sigma)$, and $(pp\pi)$ are Slater-Koster
parameters between $pd$ and $pp$, respectively~\cite{Harrison1999}.
The contribution of the order of $(pd\pi)^2/(\epsilon_d-\epsilon_p)$
cancel out in the honeycomb lattice, in sharp contrast to
$\mathrm{Sr_2IrO_4}$ with the perovskite
lattice~\cite{Kim2008,Kim2009}. The NNN transfer integrals are spin
dependent, and it is essential for the realization of the
topological insulator phase. For intra-plane in A plane,
$1\rightarrow1$ hopping can be written as
\begin{eqnarray}
\hat{t}^{\prime}_{11}&=&it^{\prime}_{1\parallel}\vec{\sigma}\cdot\vec{r}_{11}+t_{1\parallel},
\end{eqnarray}
$\vec{r}_{11}$ is a unit vector
$\vec{r}_{11}\propto\vec{t}_{11}+1/\sqrt{2}\hat{z}$, $\vec{t}_{11}$
is the hopping link.
$\hat{t}^{\prime}_{22}=\hat{t}^{\prime\dagger}_{11}$ due to
inversion symmetry. While in B plane,
$\hat{t}^{\prime}_{33}=e^{-i\pi/2\sigma_z}\hat{t}^{\prime}_{11}e^{i\pi/2\sigma_z}$,
$\hat{t}^{\prime}_{44}=e^{-i\pi/2\sigma_z}\hat{t}^{\prime}_{22}e^{i\pi/2\sigma_z}$
due to $C_2$ symmetry. For intra-plane ($A\rightarrow B$),
\begin{equation}
\hat{t}^{\prime}_{13}=it_{2\perp}\vec{\sigma}\cdot\vec{r}_{13}+t_{2\perp}
\end{equation}
$\vec{r}_{13}$ is a unit vector
$\vec{r}_{13}\propto\vec{t}_{13}-\alpha\hat{z}$, $\vec{t}_{13}$ is
the hopping link, $\alpha$ is some parameter which depend on
materials and cannot be determined purely by symmetry, below we
choose $\alpha=1/\sqrt{2}$ which has almost the same amplitude as
intra-plane. $\hat{t}^{\prime}_{24}=\hat{t}^{\prime}_{13}$,
$\hat{t}^{\prime}_{14}=\hat{t}^{\prime}_{23}=e^{-i\pi/6\sigma_z}\hat{t}^{\prime}_{13}e^{i\pi/6\sigma_z}$.
Explicitly, $\vec{r}_{ij}$ for the intra-plane
$1\rightarrow1,2\rightarrow2$ hopping are $x_1,y_1,z_1$, and
$3\rightarrow3,4\rightarrow4$ are $x_2,y_2,z_2$ denoting in
Fig.~\ref{fig1}(a).

\begin{figure}[t]
\centering
\includegraphics[width=3.0in,clip=true]{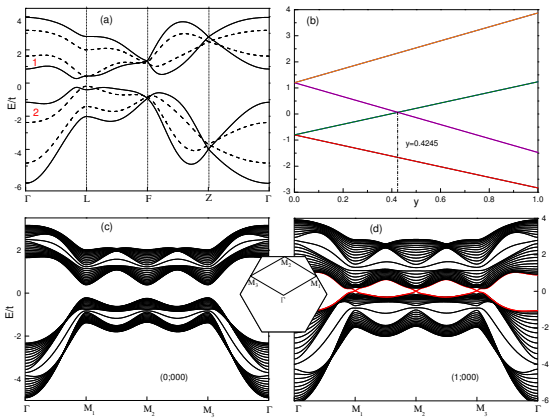}
\caption{(color online) (a) 3D energy band dispersion of
tight-binding model for corundum structure with $\lambda=0.8$, and
$y=0.3$(dashed line), $y=0.55$(solid line). (b) The change of energy
levels at $F$ point $(\pi00)$ versus $y$. A band crossing occurs at
$y=0.4245$. The system changes from trivial insulator to topological
insulator. (c) \& (d) 2D band structure for a slab with 001 surface
for the parameters $\lambda=0.8$, $y=0.3$ in (c) and $y=0.55$ in
(d). The red curves in (d) stands for surface states. The inset
shows the surface Brillouin zone.}
\label{fig2}
\end{figure}

In summary, the transfer integrals are real and spin independent for
NN links, while complex and spin dependent for NNN links. The
accurate hopping parameters varies in different materials. As an
example, in the following we will use the transfer integrals of Ir
oxide introduced in Ref.~\cite{Nagaosa2009}. One can always define
all the parameters in the unit of in-plane nearest neighbor hopping
$t$, which leads to $t=1$, $t^{\prime}_{1\parallel}=0.33$,
$t_{1\parallel}=-0.1$, $t_{\perp}=y$, $t_{2\perp}=0.5y$,
$t^{\prime}_{2\perp}=\lambda t_{2\perp}=0.4y$. Here $\lambda$ is the SOC strength which determines the
ratio of spin-dependent hopping and spin-independent hopping. For Ir
oxide we have $\lambda=0.8$. All the inter-plane hopping matrix
elements are rescaled by a factor $y$ which incorporates the
anisotropy between intra-plane and inter-plane directions. The
energy dispersion for $y=0.3$ (dashed line) and $y=0.55$ (solid
line) are shown in Fig.~\ref{fig2}(a), which shows that the system
at half filling is an insulator in both case. Due to inversion
symmetry, all the energy bands are doubly degenerate.

\begin{figure}[t]
\centering
\includegraphics[width=2.5in,clip=true]{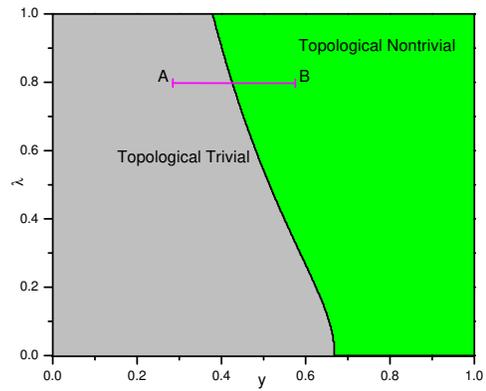}
\caption{(color online) The phase diagram of the system with two
variables: the out of plane hopping parameter $y$ and SOC parameter
$\lambda$. The green and gray regions stand for topological
nontrivial and trivial phases, respectively. Point A and B
correspond to the parameters used in Fig.~\ref{fig2}(c) and (d),
respectively.} \label{fig3}
\end{figure}

In three-dimensional topological band insulators, four independent
$Z_2$ topological invariants can be
defined~\cite{moore2007,fu2007b,roy2009}. For inversion symmetric
systems, all the topological invariants can be simply determined by
the parity of the wave-functions at the 8 time-reversal invariant
momenta (TRIM) in the Brillouin zone~\cite{fu2007a}. Denote ${\bf
G}_1,~{\bf G}_2,~{\bf G}_3$ are the three basis vectors of the
reciprocal lattice, then the 8 TRIM's are defined by ${\bf
k}_i=(k_{1}{\bf G}_1+k_{2}{\bf G}_2+k_3{\bf G}_3)/2\pi$ with $k_1,k_2,k_3=0$ or %Jing, what's the standard definition of reciprocal lattice vector? Shall we use 1/2 or pi here. Please make the
%whole draft consistent
$\pi$. For each TRIM ${\bf k}_i$, one can define a $Z_2$ quantity
$\delta_i$ as the multiplication of the parity of all occupied bands
$\delta_i=\prod_{s\in{\rm occ}}\xi_s$, with $\xi_s$ the parity of
$s$-th band. It should be noticed that a Kramers pair of bands are
only counted once, otherwise $\delta_i$ would always be even. The
four $Z_2$ invariants $(\nu_0;\nu_1\nu_2\nu_3)$ can be determined by
$\delta_i$ in which $\nu_0=\prod_{i=1}^8\delta_i$ is the strong
topological invariant which is stable upon disorder, and responsible
for the topological magneto-electric effect\cite{qi2008}. More
discussions on the other three "weak topological invariants" can be
found in Ref.\cite{fu2007a}. For $y=0.3$, we find $\delta=+1$ at all
TRIM at half filling. On the contrary, for $y=0.55$ we find
$\delta=-1$ at the three $F$ points (see Fig. \ref{fig1} (b)) and
$\delta=+1$ at all other TRIM's. Consequently, $y=0.55$ phase is a
strong topological insulator(STI) with the topological character
$(1;000)$, and $y=0.3$ is a trivial insulator with character
$(0;000)$. From this result we see that a band
inversion\cite{bernevig2006d} occurs at $F$ points upon the change
of $y$. In Fig. \ref{fig2} (b) we show the energy at $F$ point
versus $y$, from which one can see clearly a level crossing at
$y\simeq 0.42$. The topological invariants can be calculated for all
values of anisotropy parameter $y$ and spin-orbit coupling parameter
$\lambda$, which leads to the phase diagram shown in Fig.
\ref{fig3}. One can see that the topological nontrivial band
structure can be realized at large $y$ (i.e., small anisotropy) even
for {\em infinitesimal} spin-orbital coupling. However, one should
notice that for some parameters the band structure is actually a
semi-metal (similar to $Sb$), which has a direct gap but
does not have in-direct gap. We also solve the Hamiltonian (\ref{H})
in a slab geometry with two 001 surfaces to study explicitly the
topological surface states. Fig.~\ref{fig2}(c)\&(d) shows the 2D
energy dispersion of the two systems shown in Fig.~\ref{fig2}(a).
In addition to the bulk states, for $y=0.55$ there are surface
states with three Dirac cones at $M$ points of the surface BZ, while
no surface state is found for $y=0.3$, in consistency with the bulk
topological invariants.

To get better understanding of the physical properties of this
system, a low energy effective model can be obtained by expanding
the Hamiltonian around the $F$ points. Around each $F$ point, the
effective model is $4\times 4$ which describes two Kramers pairs of
low lying bands and has Dirac-like form. In the following, we will
denote the momentum by its coordinate in the basis of reciprocal
lattice, {\it i.e.}, ${\bf k}=\left(k_1{\bf G}_1+k_2{\bf
G}_2+k_3{\bf G}_3\right)/2\pi$. The $F$ points are given by
$(\pi,0,0),~(0,\pi,0)$ and $(\pi,\pi,0)$. Around the point
$(\pi,0,0)$ the Hamiltonian has the following form:
\begin{equation}
\mathcal{H}_{\mathrm{eff}}(\pi00)=\epsilon_0(\mathbf{q})\mathbb{I}_{4\times4}+\sum_{a=1}^5
d_a(\mathbf{q})\Gamma_a,\label{effectiveHamiltonian}
\end{equation}
Here the Dirac $\Gamma$-matrices are defined as
$\Gamma_a=(\tau_x\otimes\sigma_x,\tau_x\otimes\sigma_y,\tau_y\otimes1,\tau_z\otimes1,\tau_x\otimes\sigma_z)$
where $\tau_i$ and $\sigma_i$ ($i=x,y,z$) denote the Pauli matrices
in the space of orbital and spin, respectively.
$\mathbf{q}=\mathbf{k}-(\pi00)$,
$d_a(\mathbf{q})=\sum_{i=1,2,3}A^a_iq_i$ for $a=1,2,3,5$,
$d_4(\mathbf{q})=M+\sum_{i=1,2,3}B_iq_i^2$, and
$\epsilon_0(\mathbf{q})=C+\sum_{i=1,2,3}D_iq_i^2$. For
$\lambda=0.8$, around the topological phase transition point we have
\begin{eqnarray}
A_i^a&=&\left(\begin{array}{cccc}0.14&-0.12&0.37&-0.34\\-0.47&0.06&-0.13&0.09\\0.014&0.038&0.015&0.055\end{array}\right),\nonumber\\
B_i&=&\left(0.625, 0.32, 0.24\right),\nonumber\\
D_i&=&\left(0.375, 0.04, 0.04\right)\nonumber
\end{eqnarray}
$C=0.064$. The mass parameter $M$ depends on $y$ as
$M\approx-23y+9.76$ which changes sign at $y\simeq 0.42$ and leads
to the topological phase transition. The effective Hamiltonian
around the other two $F$ points at $(0\pi0)$ and $(\pi\pi0)$ can be
obtained by $C_3$ rotation.

\begin{figure}[t]
\centering
\includegraphics[width=2.5in, clip=true]{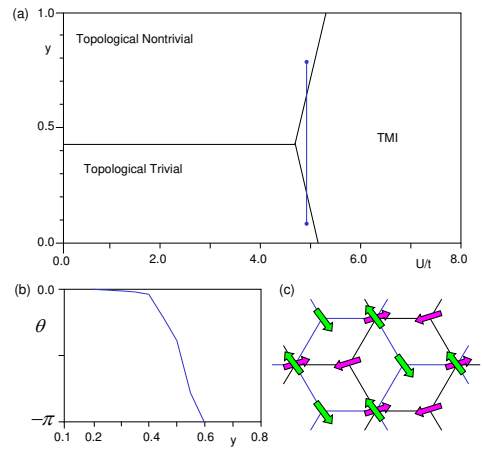}
\caption{ (Color Online) (a) The phase diagram with the out of plane
hopping $y$ and onsite repulsion $U$ as parameters. The phase on the
right is the topological magnetic insulator (TMI) which carries the
dynamic axion. (b) The value of $\theta$ (see text) along the blue
line in the phase diagram. (c) The SDW order pattern. The purple
arrows represent the spin in the honeycomb layer A and the green
arrows represent the spin in the adjacent honeycomb layer B. Other
possible spin configurations can be obtained by six-fold rotations
of this one.}\label{fig4}
\end{figure}

Now we study the effect of electron correlation. The leading term in
the interaction Hamiltonian is the onsite Hubbard repulsion for the
$j_{\rm eff}=1/2$ orbitals
\begin{equation}
H_{\mathrm{int}}=U\sum\limits_{i}n_{i\uparrow}n_{i\downarrow},
\end{equation}
Magnetic ordering in this system can be studied in meanfield
approximation. For simplicity, we will only consider the order
parameters that do not break translation symmetry. The mean-field
calculation predicts a SDW phase above a critical $U$, as shown in
the phase diagram in Fig. \ref{fig4}. The spin moments of this
SDW phase lie in the honeycomb plane, which are ordered
antiferromagnetically within each honeycomb layer and non-collinear
between the two neighboring honeycomb layers, as shown in Fig.
\ref{fig4}.

Such a SDW order breaks both time-reversal symmetry $\mathcal{T}$
and inversion symmetry $\mathcal{P}$ spontaneously, but preserves
the combination $\mathcal{P}\mathcal{T}$. Thus the magneto-electric
coefficient $\theta$ in the $\theta {\bf E\cdot B}$ term may deviate
from the time-reversal invariant values $0$ and $\pi$ in the SDW
phase, and certain spin wave fluctuations are coupled to photon as
axions\cite{Li2010}. To estimate the value of $\theta$ in the SDW
phase, we can study the effect of the SDW order in the effective
model (\ref{effectiveHamiltonian}). To the leading order, the
$\mathcal{T}$, $\mathcal{P}$ breaking but $\mathcal{T}\mathcal{P}$
preserving perturbation to the effective model must have the form
$\delta{H}_{\mathrm{eff}}(\mathbf{q})=\sum_{a=1,2,3,5}m_a\Gamma_a$,
where $m_a$ depends linearly on the spin moment $\left\langle {\bf
S}_i\right\rangle$ obtained from mean-field theory. Thus the
perturbed Hamiltonian can still be written as
$\mathcal{H}(\mathbf{q})=\sum_{a=1}^5d_a(\mathbf{q})\Gamma_a$ with
$d_a(\mathbf{q})(a=1,2,3,5)=\sum_{i=x,y,z}A^a_ik_i+m_a$ and
$d_4({\bf q})$ unchanged. For this effective model, $\theta$ has an
explicit expression~\cite{Li2010}
\begin{equation}
\theta=\frac{1}{4\pi}\int
d^3k\frac{2|d|+d_4}{(|d|+d_4)^2|d|^3}\epsilon^{ijkl}d_i\partial_xd_j\partial_yd_k\partial_zd_l
\end{equation}
where $i,j,k,l=1,2,3,5$, and $|d|=\sqrt{\sum_{a=1}^5d_a^2}$. Since
the main contribution to $\theta$ comes from the region close to
Dirac points, $\theta$ can be approximated by the sum of $\theta$s
calculated separately for each Dirac point using the effective
model. The numerical results of $\theta$ is shown in Fig. \ref{fig4} (b).

We wish to thank T. L. Hughes, Ian Fisher, Z. J. Xu and B. F. Zhu
for insightful discussion. This work is supported by the NSF under grant numbers DMR-0904264 and by the Keck Foundation. JW
acknowledges the support of China Scholarship Council, NSF of China
(Grant No.10774086), and the Program of Basic Research Development
of China (Grant No. 2006CB921500).

\end{document}